# Metal passivation effect on focused beam-induced nonuniform structure changes of amorphous SiO$_x$ nanowire

Jiangbin Su[1,2], Xianfang Zhu[1,3*], Liang Cheng[1]

1. China-Australia Joint Laboratory for Functional Nanomaterials & Physics Department, Xiamen University, Xiamen 361005, People's Republic of China

2. Experiment Center of Electronic Science and Technology, School of Mathematics and Physics, Changzhou University, Changzhou 213164, People's Republic of China

3. Institute of Biomimetics and Soft Matter, Xiamen University, Xiamen 361005, People's Republic of China

* Correspondence and request for materials should be addressed to X.F.Z. (email: zhux@xmu.edu.cn)

**Abstract**

Passivation effect of heterogeneous Au nanoparticles (AuNPs) on the nonuniform structure changes of amorphous SiO$_x$ nanowire (a-SiO$_x$ NW) as athermally induced by focused electron beam (e-beam) irradiation is investigated in an *in-situ* transmission electron microscope. It is found that at room temperature the straight and uniform a-SiO$_x$ NW demonstrates an accelerated necking at the nanoscale along with a fast, plastic elongation and a local S-type deformation in the axial direction. However, once being modified with uniform AuNPs, the nanocurved sidewall surface of a-SiO$_x$ NW becomes intriguingly passivated and the processing transfers from a diffusion (or plastic flow)-dominated status to an evaporation (or ablation)-dominated status. As a result, the necking of the AuNPs-modified a-SiO$_x$ NW is greatly retarded without visible elongation and bending deformation. Two combined effects of nanocurvature and beam-induced soft mode and instability of atomic vibration are further proposed to elucidate the observed new phenomena.

# 1. Introduction

Due to the quasi-one dimensional wire shape and the amorphous structure, $SiO_x$ nanowires (NWs) are thermodynamically structure unstable. It has been reported that under energetic electron beam (e-beam) irradiation, the amorphous $SiO_x$ (a-$SiO_x$) NWs can demonstrate structural changes or transformations kinetically such as S-type deformation, elongation, cutting, necking or shrinking [1-2]. Among these, the focused beam-induced necking of a-$SiO_x$ NW accelerated at the nanoscale and accompanied by a local S-type deformation and plastic elongation [1] is especially intriguing and interested. It not only demonstrates the atomic vibration soft mode and instability effect [3-4] as athermally induced by focused e-beam irradiation but also evidences the intrinsic nanocurvature effect [4-5] of a-$SiO_x$ NW experimentally. Further, it has been reported that the above two effects are much more obvious in an amorphous low dimensional nanostructure (LDN) than in a crystalline one [1]. Hence, it can be predicted that heterogeneous crystalline nanoparticles (NPs) such as metal NPs decorated on the surface of an amorphous Si-O NW may passivate the beam-induced structural changes such as the aforementioned local S-type deformation, elongation and necking. However, there is still no experimental evidence to demonstrate such passivation effect of metal NPs on the S-type deformation, elongation and necking of amorphous NW and thus the underlying physical mechanism remains unclear. In this regard, study on the passivation effect of metal NPs on amorphous NW is imperative and crucial not only to the precise and flexible controlling of focused beam-induced NW structural changes or processing but also to the fundamental understanding of some new phenomena and concepts in LDNs.

## 2. Experimental

The well-defined, straight a-SiO$_x$ NWs with smooth surface and uniform diameter of about 45 nm were grown by our improved chemical vapor deposition technique where x is determined to be 2.3 [6]. Some uniform, dispersive AuNPs were further deposited onto the a-SiO$_x$ NWs surface for a total duration of about 3 s by a cool sputter coater (BAL-TEC SCD 005). Both the pristine and the modified a-SiO$_x$ NWs were ultrasonically detangled in ethanol and then dripped onto the holey carbon film on Cu grids to prepare the TEM specimens. The e-beam irradiation was carried out at room temperature in a field-emission Tecnai F-30 TEM operated at 300 kV. The current density of e-beam was kept at about $3.6 \times 10^2$ A/cm$^2$ (flux: $3.5 \times 10^5$ nm$^{-2}$ s$^{-1}$) with its diameter of about 60 nm, which was a little larger than that of the wires before irradiation. The focused beam was always targeted on the radial center of single wire segments protruding into the open space of the holes in the carbon film of microscopy grid. In each frame, the radius of the wire (half of the diameter) was taken as the average value measured at the thinnest location of the irradiated segment; the length and the volume of the wire segment were taken as the average values measured and calculated between two red mark dots in Figs. 1-2 along the wire axis, which were carefully marked at the locations of two feature points and by further checking the relative positions of their surrounding feature points on the wire surface (see the arrows in Figs. 1-2).

## 3. Results and discussion

The sequential TEM micrographs in Figs. 1-2 show the comparative structure evolutions between a pristine (without AuNPs) and a modified (with AuNPs) a-SiO$_x$ NW segment during the same irradiation of a focused e-beam at room temperature. Before irradiation, as shown in (a) of Figs. 1-2, both the pristine and the modified a-SiO$_x$ NW segments presented a well-defined, straight wire shape with nearly the same initial diameter of ~45 nm but different surface morphologies. For the pristine NW, the surface is smooth and clean (see Fig. 1(a)); while for the modified NW, the surface is uniformly decorated by many tiny, dispersive AuNPs with an average size of ~3.4 nm (see Fig. 2(a)). With the focused irradiation turned on, as demonstrated in Fig. 1, the pristine NW segment within the irradiated area and nearby demonstrates a rapid, local necking which seems to be accelerated when the minimum radius at the center of the beam reduces down to about 2.3 nm (see Fig. 3(a)). It eventually leads to a quick breakage at the center of the beam after 26 s irradiation (see Fig. 1(f)). Accompanied with the rapid necking, a fast, plastic elongation and S-type deformation in the axial direction of wire is also occuring resultingly (see Figs. 1 and/or 3(b)). In contrast, as shown in Fig. 2, the modified NW segment seems much more stable and demonstrates a slower necking which further slows down with the increase of irradiation time (see Fig. 3(a)). Throughout the slow necking, there is no visible elongation and bending deformation (i.e. keep straight) in the axial direction consequently (see Figs. 2 and/or 3(b)). Furthermore, the crystalline AuNPs on the wire surface seem to be much more stable and rarely evaporate even after the surrounding a-SiO$_x$ material is gradually polished or dug out. Instead, they tend to get close to each other, then aggregate together and finally form a single crystal Au nanobridge within the irradiated area (see Figs. 2 and 4). At the same time, the a-SiO$_x$ materials are totally evaporated and lost especially at the center of the beam (see Fig. 2(h-i)). The comparative findings indicate that there is a notable passivation effect of AuNPs on the focused beam-induced nonuniform structural changes of a-SiO$_x$ NW. Meanwhile, as shown in Fig. 3(c), there is almost no volume loss observed in the pristine NW segment throughout the irradiation whereas the volume of the modified NW segment decreases continuously with the irradiation time. It indicates that the existence of AuNPs on the wire surface or the passivation effect of AuNPs greatly influences the transportation modes of material atoms such as atom diffusion (or plastic flow) and atom evaporation (or ablation) [1-2]. Similar irradiation on other pristine and AuNPs-modified NW segments is repeated several times. It is observed that the features of the structural changes are essentially the same as shown in Figs. 1-3.

The above comparative observations strongly demonstrate a notable passivation effect of AuNPs on the structural changes of a-SiO$_x$ NW. Obviously, the as-induced nanophenomena can not be fully explained by the existing knock-on mechanism or simulations [7-11]. This is because the knock-on mechanism and simulations are at the first place built on consideration of the nature of equilibrium, symmetry, periodicity, and linearity of bulk crystalline structures or its approximation whereas the beam-induced nanophenomena are intrinsically of non-equilibrium, amorphous, and non-linear nature. The previous research [1-2,4,12-14] has demonstrated that our proposed nanocurvature effect [4-5] and beam-induced atomic vibration soft mode and instability effect [3-4] are universal concepts and can well explain the beam-induced nanophenomena. From the following analysis, we can conclude that the nanocurvature effect of the a-SiO$_x$ NW and the e-beam-induced atomic vibration soft mode and instability effect are the two key factors to induce the observed comparasive structural changes of the a-SiO$_x$ NWs.

As discussed in literature [1-2], when the radius of a NW approaches its atomic bond length, a positive nanocurvature on the highly curved wire surface will become appreciable. Such a positive nanocurvature would cause an additional tensile stress on the electron cloud structure of surface atoms which could lead to a dramatic increase in the wire surface energy. This dramatically increased surface energy would give rise to a strong tendency of self-compression on the nanocurved wire surface and thus provide a thermodynamic driving force for the wire to contract in the radius direction and massive atom plastic flow as self-extruded towards the two ends of the wire. These phenomena are so-called nanocurvature effect or nanosize effect in a broad sense [4-5]. It is also applicable to other LDNs such as NPs. However, the surface structure of the crystalline AuNPs of metal bond tends to be atom-faceted. This atom-faceted surface structure probably could avoid the deformation of the electron cloud of surface atoms [4,15] and thus lower down the surface strain energy. In contrast, the sidewall surface of the a-SiO$_x$ NW of Si-O-Si bridge bond has no such a atom-faceted surface structure. Thus, the electron cloud on the sidewall surface of the a-SiO$_x$ NW should be highly deformed and thus induces high surface strain energy by nanocurvature when the surface is curved to the nanoscale. Accordingly, we can conceive that the surface energy of the AuNPs should be lower than that of a-SiO$_x$ NW if they are all curved to the same nanoscale or even the AuNPs are much more curved. The structure differences probably lead to the fact that the nanocurvature effect of the AuNPs is much weaker than that of a-SiO$_x$ NW. In this way, the AuNPs modified on the a-SiO$_x$ NW surface become the most stable sites and tend to keep their crystalline structure and thus their passivation effect throughout the irradiation.

Although the a-SiO$_x$ NW is thermodynamically instable, only the nanocurvature effect is not enough to cause the structural changes kinetically for the a-SiO$_x$ NW at room temperature. For the transition from one metastable structure configuration to another, the a-SiO$_x$ NW needs to be softened or overcome some energy barrier. With the assistance of energetic e-beam-induced athermal activation, it can be reasonably assumed that under the highly energetic e-beam irradiation the energy deposition rate [2] is so fast that it can be comparable with frequencies of thermal vibration of atoms in condensed matter. Thus, there would be no enough time for the atoms to transfer the beam-deposited energy into atom vibration energy and the mode of atom thermal vibration would become softened or the vibration of atoms would lose stability. The as-induced atomic vibration soft mode and instability would suppress the energy barrier greatly or even make it totally disappear. In doing so, the irradiation would induce athermal diffusion (even athermal plastic flow) or/and athermal evaporation (even athermal ablation) of the wire atoms. These phenomena are so-called beam-induced atomic vibration soft mode and instability effect, beam-induced athermal activation effect or nanotime effect in a broad sense [3-4]. Relative to the amorphous $SiO_x$ NW, the crystalline AuNPs of metal bond would hardly become amorphized by e-beam irradiation and thus are much more resistant to the irradiation with their metal crystal structure (see Fig. 4). As a result, the beam-induced atomic vibration soft mode and instability effect of AuNPs is much weaker than that of a-$SiO_x$ NW and a notable passivation effect of AuNPs on the beam-induced structure changes of a-$SiO_x$ NW will exist when the surface of a-$SiO_x$ NW is decorated by AuNPs.

In this way, for the case of the pristine a-$SiO_x$ NW as illustrated in Fig. 5(a), the irradiated wire segment is modulus-softened under energetic e-beam irradiation. Due to the nanocurvatue-caused strong tendency of self-compression on the nanocurved wire surface, the softened wire segment necks accordingly in the radius direction and prolongs in the axial direction by plastic flow of massive atoms quickly. The quick elongation extending to the two ends of the wire would be somewhat hindered by the supporting carbon film and thus causes the S-type deformation. In particular, with the increase of irradiation time, the NW became thinner especaill at the center of the beam and demonstrated an accelerated radial shrinkage at the nanoscale as shown in Fig. 3(a). This could be attributed to the dramatically increased surface energy or nanocurvature effect of the NW when it became thinner with a radius down to the nanoscale. In contrast, for the case of the AuNPs-modified a-$SiO_x$ NW as illustrated in Fig. 5(b), the AuNPs greatly reduce the surface energy of the a-$SiO_x$ NW and thus passivate the nanocurved wire sidewall surface. That is, the AuNPs could greatly weaken or even block the diffusion and plastic flow of atoms of a-SiO$_x$ NW. Instead, the wire atoms tend to evaporate and escape from the wire surface especially at the center of the beam (i.e. preferential evaporation relative to that at the edge of the beam, see Fig. 5(b)), which has a Gaussian-like intensity profile on specimen with a higher flux or beam energy deposition rate. As a consequence, as illustrated in Fig. 5(b), the modified a-SiO$_x$ NW necks slowly by atom evaporation without axial elongation and keep the wire axis straight throughtout the irradiation.

For the a-SiO$_x$ NW without AuNPs as shown in Figs. 1 and 5(a), it can be regarded as an amorphous cylinder-like structure of Si-O-Si bridge bond. After being modified with AuNPs as shown in Figs. 2 and 5(b), part of the smooth, amorphous sidewall surface of the a-SiO$_x$ NW is uniformly replaced by the atom-faceted, crystalline AuNPs surface. At this stage, it is still hard to conceive exactly how the metal-bonded Au atoms interact with the bridge-bonded Si-O-Si of the nanocurved wire sidewall. Nevertheless, it has demonstrated an extraordinary structural stability of the crystalline AuNPs under e-beam irradiation and their enhancement in the stability of the highly curved a-SiO$_x$ NW. We also note that although the AuNPs are hardly evaporated and amorphized, they would get close to each other, aggregate together and finally turn into a single crystal Au nanobridge within the irradiated area (see Figs. 2 and 5). It is expected that the movement or position adjustment of the AuNPs is mainly caused by the gradual evaporation of the surrounding SiO$_x$ material, which leads to a reducing of wire diameter and sidewall surface area for the AuNPs to "sit" on. In the process, the two types of nanoripening of the AuNPs would take place, leading to a reduction in total surface energy, which were driven by the non uniformly-distributed nanocurvature over surface of the AuNPs. That is, during the movement, the larger AuNPs may grow by taking up of the adjacent smaller ones. Or, when two or more AuNPs contact together, they coalesce and grow into a biger AuNP (see Fig. 2) probably accompanied by the rotation of AuNPs or the adjustment of crystal orientations. As the irradiation goes on, more and more AuNPs participate in the growing of the big AuNP which finally leads to the formation of a Au nanobridge (see Fig. 2). Although the detailed observation and mechanism are not found, the nanoripening of the AuNPs will be reported in a separated work (Zhu XF et al, to be submitted).

In addition, for the modes of mass transportation, the atom diffusion and plastic flow of massive atoms would lead to a re-distribution of wire atoms without volume loss whereas the atom evaporation and ablation would cause the atoms to escape from the wire surface with a pure volume loss. According to the comparative results of volume change as shown in Fig. 3(c), we can conclude that the processing transfers from a diffusion (or plastic flow)-dominated status to an evaporation (or ablation)-dominated status after the surface of a-SiO$_x$ NW is modified by uniform AuNPs. It demonstrates that the existence of AuNPs or their passivation effect would greatly affect or even determine the modes of beam-induced mass transportation of a-SiO$_x$ NW.

## 4. Summary

In this work, the passivation effect of heterogeneous AuNPs on the nonuniform structure changes of a-SiO$_x$ NW as athermally induced by focused e-beam irradiation is investigated in an *in-situ* transmission electron microscope. It is found that at room temperature the straight and uniform a-SiO$_x$ NW demonstrates an accelerated necking at the nanoscale accompanied with a fast, plastic elongation and a local S-type deformation in the axial direction. However, once being modified with uniform AuNPs, the nanocurved sidewall surface of a-SiO$_x$ NW becomes intriguingly passivated. Furthermore, the processing transfers from a diffusion (or plastic flow)-dominated status to an evaporation (or ablation)-dominated status after the surface of a-SiO$_x$ NW is modified by uniform AuNPs. As a consequence, the necking of the AuNPs-modified a-SiO$_x$ NW is greatly retarded without visible elongation and bending deformation. The above processes involve accelerated, surface-sensitive, athermal and ultrafast atom transportations and thus cannot be fully explained by the existing knock-on mechanism and simulations. But it can be well interpreted by a novel mechanism of athermal diffusion and plastic flow and/or athermal evaporation and ablation of wire atoms as driven by the nanocurvature of a-SiO$_x$ NW and the e-beam-induced soft mode and instability of atom vibration.

## Acknowledgements

This work was supported by the NSFC project under grant no. 11574255, the Science and Technology Plan (Cooperation) Key Project from Fujian Province Science and Technology Department under grant no. 2014I0016, and the National Key Basic Science Research Program (973 Project) under grant no. 2007CB936603.

Thesis, Australian National University, Canberra, 2000.

**Figures**

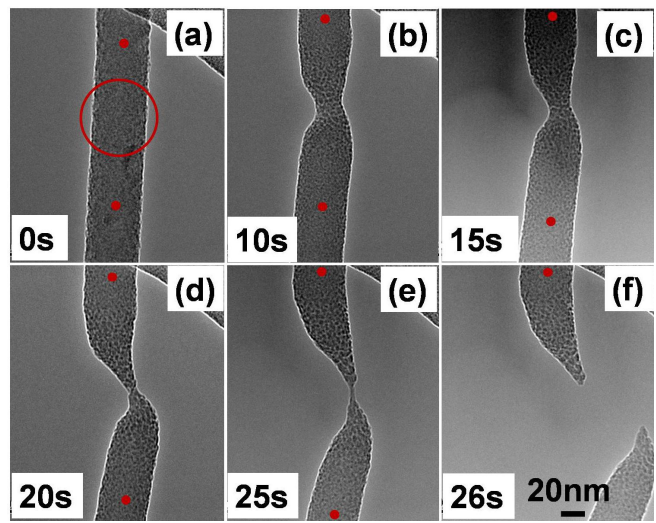

**Fig. 1** Sequential TEM images showing local structure changes with the irradiation time as induced by irradiation of a focused beam (of ~60 nm in beam diameter with current density of $3.6 \times 10^2$ A/cm$^2$) on a pristine a-SiO$_x$ NW (of ~45 nm in wire diameter). The red circle indicates the beam spot size and also the irradiated location.

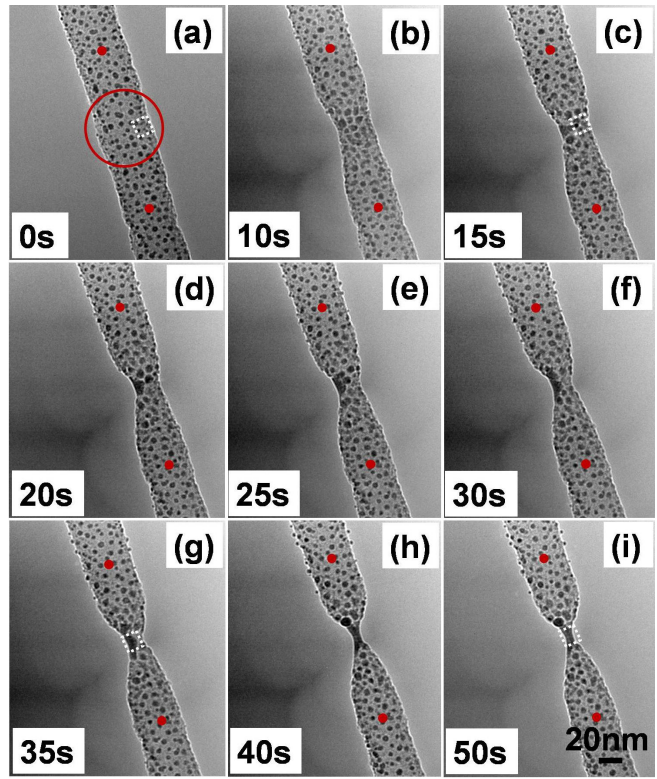

**Fig. 2** Sequential TEM images showing local structure changes with the irradiation time under the same irradiation as that in Fig. 1 on a AuNPs-modified a-SiO$_x$ NW. The red circle indicates the beam spot size and also the irradiated location.

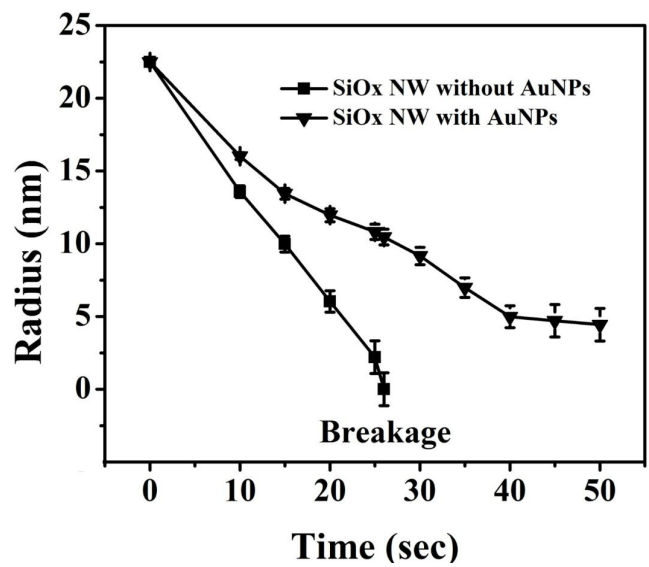

(a)

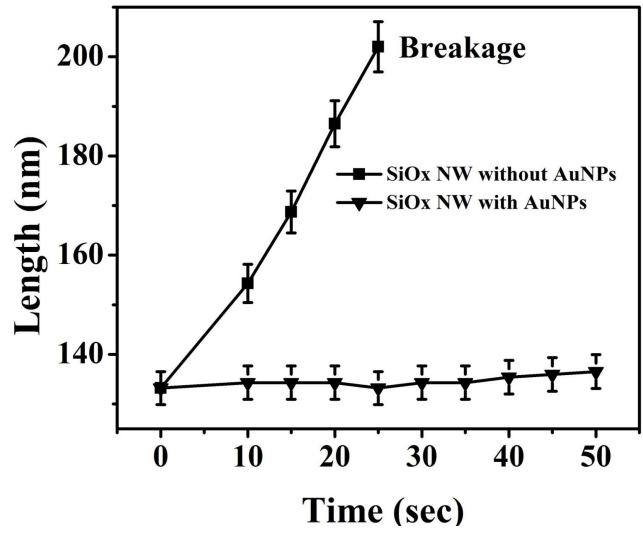

(b)

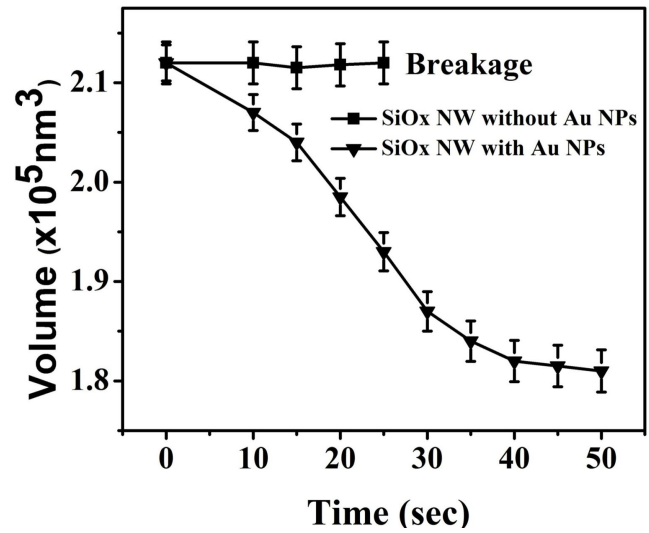

(c)

**Fig. 3** Changes of (a) radius, (b) length and (c) volume of the pristine and the modified a-SiO$_x$ NWs with the irradiation time as measured from Figs. 1-2.

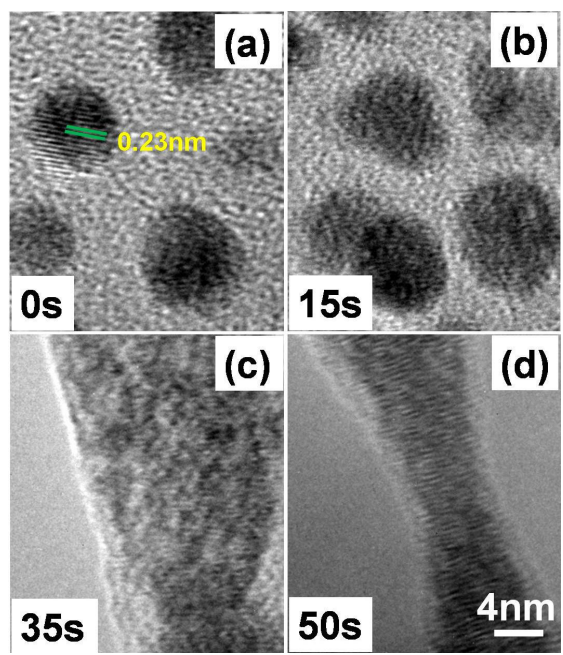

**Fig. 4** HRTEM images showing the formation processes of a single crystal Au nanobridge, which are respectively collected from the white rectangles in (a), (c), (g) and (i) of Fig. 2.

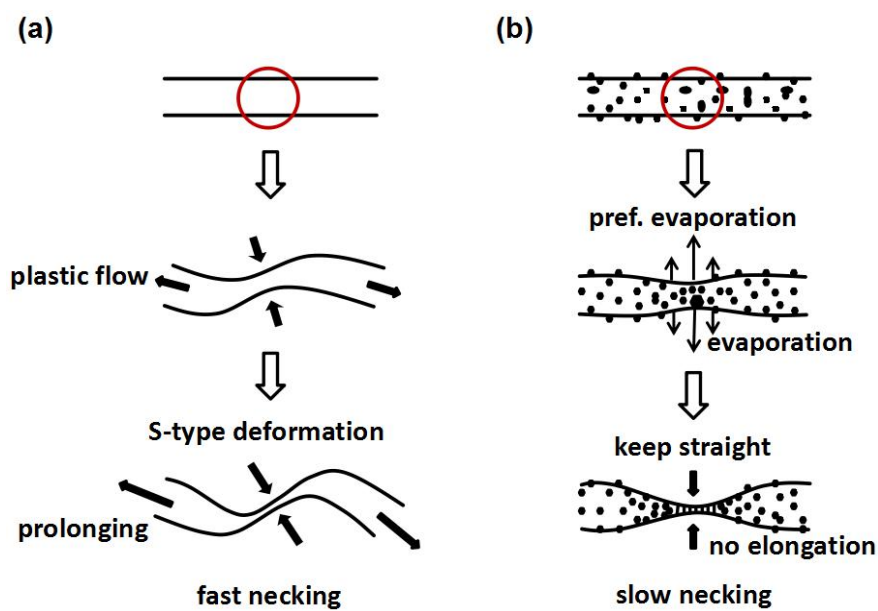

**Fig. 5** Schematic illustrations showing the passivation effect of AuNPs on the focused beam-induced nonuniform structure changes of a-SiO$_x$ NWs: (a) a-SiO$_x$ NW without AuNPs; (b) a-SiO$_x$ NW with AuNPs.